\documentclass[12pt]{iopart}
\usepackage{graphicx}
\usepackage{bm}
\usepackage{iopams}

\begin{document}
\title[Comment]{Comment on 'On the realization of quantum Fisher information'}
\author{O Olendski}\address{Department of Applied Physics and Astronomy, University of Sharjah, P.O. Box 27272, Sharjah, United Arab Emirates}
\ead{oolendski@sharjah.ac.ae}

\begin{abstract}
It is shown that calculation of the momentum Fisher information of the quasi-one-dimensional hydrogen atom recently presented by A Saha, B Talukdar and S Chatterjee (2017 \EJP {\bf 38} 025103) is wrong. A correct derivation is provided and its didactical advantages and scientific significances are highlighted.
\end{abstract}
\vskip.7in

\noindent

\submitto{\EJP}
\maketitle
Recently, a calculation of the position $I_\rho$ and momentum $I_\gamma$ Fisher informations was presented for i) a linear harmonic oscillator, ii) a quasi-one-dimensional hydrogen atom and iii) an infinite potential well \cite{Saha1}. By general definition \cite{Fisher1,Frieden1}, these quantum-information measures for arbitrary one-dimensional (1D) structures are defined as
\numparts\label{Fisher1}
\begin{eqnarray}
\label{Fisher1_X}
I_{\rho_n}&=\int\rho_n(x)\left[\frac{d}{dx}\ln\rho_n(x)\right]^2dx=\int\frac{\rho_n'(x)^2}{\rho_n(x)}\,dx\\
\label{Fisher1_K}
I_{\gamma_n}&=\int_{-\infty}^\infty\gamma_n(p)\left[\frac{d}{dp}\ln\gamma_n(p)\right]^2dp=\int_{-\infty}^\infty\frac{\gamma_n'(p)^2}{\gamma_n(p)}dp,
\end{eqnarray}
\endnumparts
where a position integration is carried out over all available interval $x$ and positive integer index $n$ counts all possible quantum states. In these equations, $\rho_n(x)$ and $\gamma_n(p)$ are position and momentum probability density, respectively:
\numparts\label{Densities1}
\begin{eqnarray}
\label{DensityX1}
\rho_n(x)&=\Psi_n^2(x)\\
\label{DensityP1}
\gamma_n(p)&=|\Phi_n(p)|^2,
\end{eqnarray}
\endnumparts
and corresponding waveforms $\Psi_n(x)$ and $\Phi_n(p)$ are related through the Fourier transformation:
\begin{equation}\label{Fourier1}
\Phi_n(p)=\frac{1}{\sqrt{2\pi\hbar}}\int\exp\!\left(-\frac{i}{\hbar}px\right)\Psi_n(x)dx.
\end{equation}
Both of them satisfy orthonormality conditions:
\numparts\label{OrthoNormality1}
\begin{eqnarray}
\label{OrthoNormalityX1}
\int\Psi_{n'}(x)\Psi_n(x)dx&=\delta_{nn'}\\
\label{OrthoNormalityP1}
\int_{-\infty}^\infty\Phi_{n'}^\ast(p)\Phi_n(p)dp&=\delta_{nn'},
\end{eqnarray}
\endnumparts
where $\delta_{nn'}=\left\{\begin{array}{cc}
1,&n=n'\\
0,&n\neq n'
\end{array}\right.$ is a Kronecker delta, $n,n'=1,2,\ldots$. Real position wave function $\Psi_n(x)$ and associated eigen energy $E_n$ are found from the 1D Schr\"{o}dinger equation:
\begin{equation}\label{Schrodinger1}
-\frac{\hbar^2}{2m_p}\frac{d^2\Psi_n(x)}{dx^2}+V(x)\Psi_n(x)=E_n\Psi_n(x),
\end{equation}
with $m_p$ being a mass of the particle and $V(x)$ being an external potential.

First, we point out that for the infinite potential well of the width $a$ the Fisher informations were calculated before \cite{LopezRosa1} where the position component was evaluated directly from Eq.~\eref{Fisher1_X} whereas for finding $I_{\gamma_n}$ an elegant and didactically instructive method was used; namely, since for this geometry both $\Psi_n(x)$ and $\Phi_n(p)$ are real, the integrand in Eq.~\eref{Fisher1_K} becomes $4[\Phi_n'(p)]^2$, and using the reciprocity between position and momentum spaces, Eq.~\eref{Fourier1}, one replaces infinite $p$ integration by the finite $x$ one:
\begin{equation}\label{IQW1}
I_{\gamma_n}=4\int_{-a/2}^{a/2}x^2\Psi_n^2(x)dx.
\end{equation}
\begin{figure}
\centering
\includegraphics[width=\columnwidth]{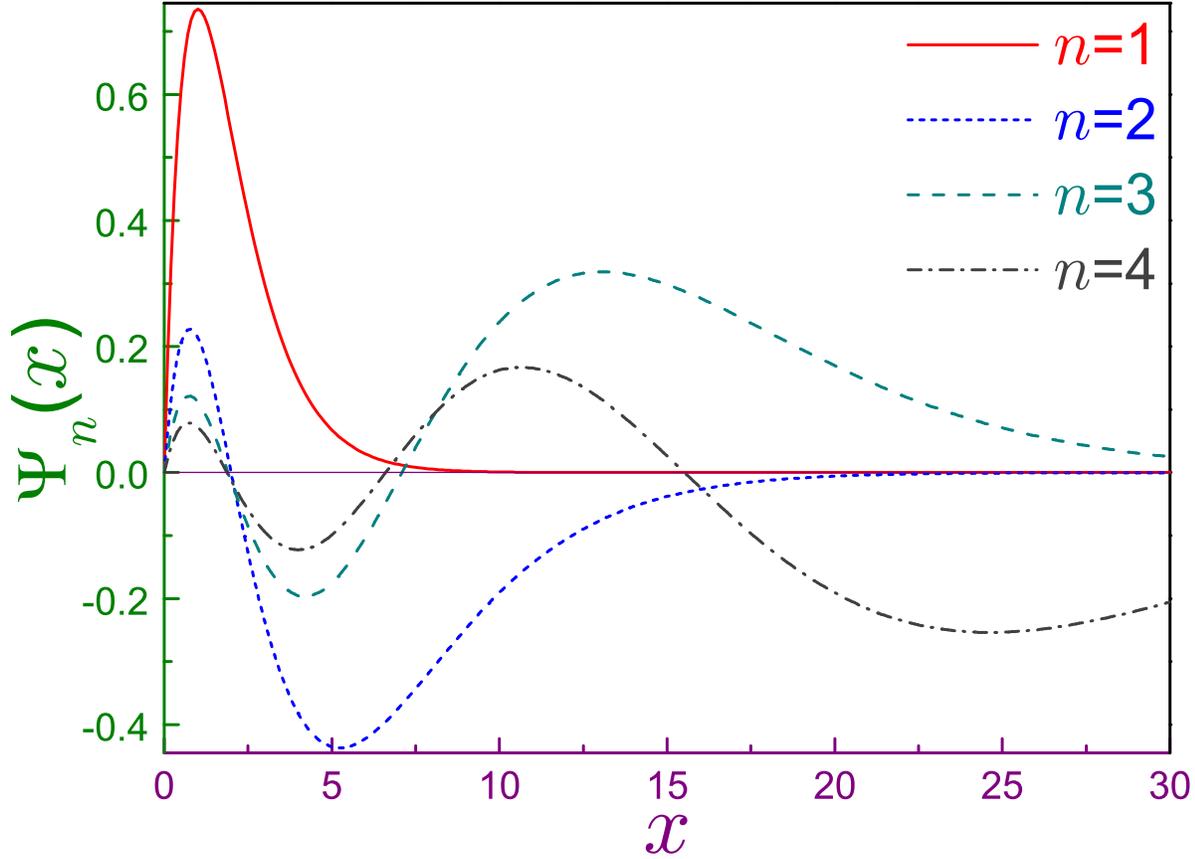}
\caption{\label{PositionFig1}
Position wave functions $\Psi_n(x)$ where solid line denotes ground state, dotted curve -- first excited level, dashed line depicts the orbital with the quantum index $n=3$, and the dash-dotted line is for the state with $n=4$.
}
\end{figure}

\begin{figure}
\centering
\includegraphics[width=\columnwidth]{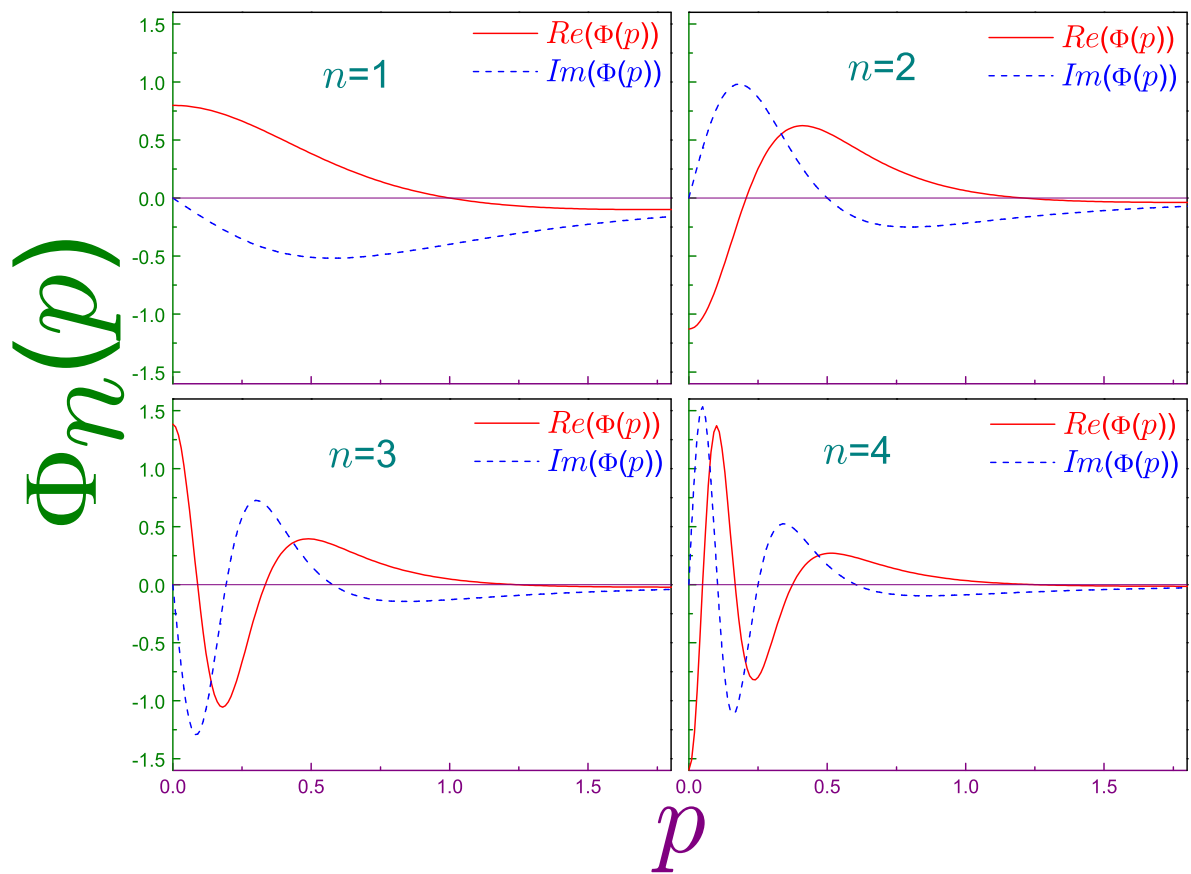}
\caption{\label{MomentumFig1}
Momentum wave functions $\Phi_n(p)$ where each panel corresponds to the quantum level with the number depicted inside it. Solid (dash) lines denote real (imaginary) components. Since real (imaginary) part of $\Phi_n(p)$ is symmetric (antisymmetric) function  of momentum, the dependencies for nonnegative $p$ only are shown.}
\end{figure}

Turning to the discussion of the quasi-1D hydrogen atom, it has to be noted that a problem of the quantum motion along the whole $x$ axis, $-\infty<x<\infty$, in the potential $V(x)\sim-|x|^{-1}$, despite its long history, still remains (owing to the strong singularity at the origin and concomitant difficulty of matching right and left solutions at $x=0$) a topic of debate and controversy, see, e.g., Refs. \cite{Palma1,Loudon1} and literature cited therein. A situation is somewhat simplified when one confines the motion only to the right half space terminated at $x=0$ by the infinite barrier. Accordingly, let us consider solutions of the Schr\"{o}dinger equation \eref{Schrodinger1} with the potential \cite{Saha1}
\begin{equation}\label{Q1Dpotential1}
V(x)=\left\{\begin{array}{cc}
-\frac{\alpha}{x},&x>0\\
\infty,&x\leq0,
\end{array}\right.
\end{equation}
$\alpha>0$. Upon introducing Coulomb units where energies and distances are measured in terms of $m_p\alpha^2/\hbar^2$ and $\hbar^2/(m_p\alpha)$, respectively \cite{Landau1}, and momenta -- in units of $m_p\alpha/\hbar$, one arrives at the differential equation
\begin{equation}\label{Equation1}
-\frac{1}{2}\frac{d^2\Psi(x)}{d^2x}-\frac{1}{x}\Psi(x)-E\Psi(x)=0,
\end{equation}
whose general solution for the negative energies, $E=-|E|$, corresponding to the bound states, reads:
\begin{eqnarray}
\Psi(x)=e^{-(2|E|)^{1/2}x}(2|E|)^{1/2}x&\left[c_1M\!\left(1-\frac{1}{(2|E|)^{1/2}},2,(2|E|)^{1/2}x\right)\right.\nonumber\\
\label{SolutionX1}
&\left.+c_2U\!\left(1-\frac{1}{(2|E|)^{1/2}},2,(2|E|)^{1/2}x\right)\right].
\end{eqnarray}
Here, $M(a,b,x)$ and $U(a,b,x)$ are Kummer, or confluent hypergeometric, functions (we follow the notation adopted in Ref.~\cite{Abramowitz1}), and $c_1$ and $c_2$ are normalization constants. Physically, this mathematical solution vanishes at the origin and, since the second item in the square brackets of the right-hand side of Eq.~\eref{SolutionX1} diverges at $x\rightarrow0$ \cite{Abramowitz1}, it has to be neglected, $c_2=0$. Remaining part must decay sufficiently fast  at infinity. From the properties of the Kummer function $M(a,b,x)$ \cite{Abramowitz1} it follows that it is possible only when its first parameter is equal to the nonpositive integer what immediately leads to the energy spectrum coinciding with the 3D hydrogen atom \cite{Landau1}
\begin{equation}\label{Energies1}
E_n=-\frac{1}{2n^2},
\end{equation}
whereas the corresponding waveform simplifies to
\begin{equation}\label{SolutionX2}
\Psi_n(x)=\frac{2x}{n^{5/2}}e^{-x/n}L_{n-1}^{(1)}\!\left(\frac{2x}{n}\right),
\end{equation}
with $L_m^{(\beta)}(x)$, $m=0,1,\ldots$, being a generalized Laguerre polynomial \cite{Abramowitz1}. Fig.~\ref{PositionFig1} shows waveforms of the first four levels. Didactically, a representation of the solution in the form of the Laguerre polynomials is much more advantageous compared to that of the confluent hypergeometric functions, Eq.~(16) in Ref.~\cite{Saha1}; in particular, utilizing properties of the Laguerre polynomials (see Eq.~2.19.14.18 in Ref.~\cite{Prudnikov2}), one instantly confirms that Eq.~\eref{SolutionX2} does satisfy the orthonormality condition, Eq.~\eref{OrthoNormalityX1}, for $n=n'$\footnote{Proof of Eq.~\eref{OrthoNormalityX1} for $n\neq n'$ is carried out in a standard way: Eq.~\eref{Equation1} for the quantum state $n$ is multiplied by $\Psi_{n'}$ and is subtracted from Eq.~\eref{Equation1} for the orbital $n'$ multiplied by $\Psi_n$ with subsequent integration.}. Moreover, the form of solution from Eq.~\eref{SolutionX2} allows an instructive calculation from Eq.~\eref{Fourier1} of the momentum waveform. For doing this, one recalls Rodrigues formula for Laguerre polynomials \cite{Abramowitz1}:
\begin{equation}\label{Rodrigues1}
L_m^{(\beta)}(x)=\frac{x^{-\beta}}{m!}\,e^x\frac{d^m}{dx^m}\left(e^{-x}x^{m+\beta}\right).
\end{equation}
Then, $\Phi_n(p)$ becomes:
\begin{equation}\label{DerivationPhiStep1}
\Phi_n(p)=\frac{1}{2\sqrt{2\pi}}\frac{n^{1/2}}{n!}\int_0^\infty e^{(1-inp)x/2}\frac{d^{n-1}}{dx^{n-1}}\left(e^{-x}x^n\right)dx.
\end{equation}
Successive $(n-1)$ integrations by parts simplify this to:
\begin{equation}\label{DerivationPhiStep2}
\Phi_n(p)=\frac{(-1)^{n-1}}{2^n\sqrt{2\pi}}\frac{n^{1/2}}{n!}\,(1-inp)^{n-1}\int_0^\infty x^ne^{-(1+inp)x/2}dx.
\end{equation}
An elementary deformation $z=\frac{1}{2}(1+inp)x$ of the integration contour in this equation yields ultimately:
\begin{equation}\label{SolutionP1}
\Phi_n(p)=(-1)^{n+1}\sqrt{\frac{2n}{\pi}}\frac{(1-inp)^{n-1}}{(1+inp)^{n+1}}.
\end{equation}
Observe that this {\em complex} solution is completely different from the {\em real} one provided by Eq.~(17) from Ref.~\cite{Saha1}. With the help of residue theorem applied to calculation of the integrals with infinite limits \cite{Arfken1}, it is elementary to check that the set from Eq.~\eref{SolutionP1} does obey Eq.~\eref{OrthoNormalityP1}, as expected. The dependencies of the real and imaginary parts of the waveforms $\Phi_n(p)$ on the momentum are shown in Fig.~\ref{MomentumFig1}. It is seen that the number and amplitude of the oscillations increase for the higher quantum indices $n$.

Knowledge of the functions $\Psi_n(x)$ and $\Phi_n(p)$ and, accordingly, of the corresponding densities
\numparts\label{Density2}
\begin{eqnarray}\label{DensityX2}
\rho_n(x)=\frac{1}{n^3}\left(\frac{2x}{n}\right)^2e^{-2x/n}L_{n-1}^{(1)}\!\left(\frac{2x}{n}\right)^2\\
\label{DensityP2}
\gamma_n(p)=\frac{2n}{\pi}\frac{1}{(1+n^2p^2)^2},
\end{eqnarray}
\endnumparts
paves the way to calculating the associated Fisher informations. Dropping quite simple intermediate computations (which in the case of the position component $I_{\rho_n}$ rely on the properties of the Laguerre polynomials \cite{Abramowitz1,Prudnikov2} while for its momentum counterpart $I_{\gamma_n}$ elementary properties of the integrals of the product of the power and algebraic functions are employed), the ultimate results are given as
\numparts\label{FisherFinal1}
\begin{eqnarray}
\label{FisherFinalX1}
I_{\rho_n}=\frac{4}{n^2}=8|E_n|\\
\label{FisherFinalP1}
I_{\gamma_n}=2n^2=|E_n|^{-1},
\end{eqnarray}
\endnumparts
what leads to the index independent product
\begin{equation}
I_{\rho_n}I_{\gamma_n}=8.
\end{equation}
Note that position Fisher information whose expression, Eq.~\eref{FisherFinalX1}, does coincide with its counterpart from Ref.~\cite{Saha1} is proportional to the absolute value of energy what is its general property while the momentum component is just the inverse of $|E_n|$. As a result, the product of the two informations stays the same for all levels. This $n$ independence singles out the hydrogen atom from other two structures studied in Ref.~\cite{Saha1} for which $I_{\rho_n}I_{\gamma_n}$ is a quadratic function of the quantum index.

\end{document}